\documentclass{article}

\usepackage{ITW06_PaperLaTeXStyle}

\usepackage{IEEEtrantools}
\usepackage{graphicx}

\usepackage{bm}
\usepackage[noadjust]{cite}
\usepackage{cases}

\newtheorem{theorem}{Theorem}
\newtheorem{lemma}{Lemma}
\newtheorem{proposition}{Proposition}

\newtheorem{remark}{Remark}

\newcommand{\eqref}[1]{\textnormal{(\ref{#1})}}
\newcommand{\relvar}[2]{\buildrel {#2} \over {#1}}
\newcommand{\eqvar}[1]{\relvar{=}{#1}}

\newcommand{\levar}[1]{\relvar{\le}{#1}}

\newcommand{\eqdef}{\eqvar{\Delta}}

\newcommand{\ceil}[1]{\left\lceil {#1} \right\rceil}

\newcommand{\plimsup}{\mathrm{p}\mbox{-}\limsup}
\newcommand{\pliminf}{\mathrm{p}\mbox{-}\liminf}

\def\QEDopen{{\setlength{\fboxsep}{0pt}\setlength{\fboxrule}{0.2pt}\fbox{\rule[0pt]{0pt}{1.3ex}\rule[0pt]{1.3ex}{0pt}}}}
\newcommand{\QED}{\QEDopen}
\newenvironment{proof}[1][Proof]{\noindent\hspace{2em}{\itshape #1: }}{\hspace*{\fill}~\QED\par\endtrivlist\unskip}

\begin{document}
\headheight = 10mm
\headsep = 6mm

\itwtitle{An Enhanced Covering Lemma for Multiterminal Source Coding}


\itwauthorswithsameaddress{Shengtian Yang, Peiliang Qiu\footnotemark[1]}
{Department of Information Science \& Electronic Engineering\\
Zhejiang University\\
Hangzhou, Zhejiang 310027, China\\
{\tt \{yangshengtian, qiupl\}@zju.edu.cn}}

\markboth{}{To appear in Proc. 2006 IEEE Information Theory Workshop, October 22-26, 2006, Chengdu, China.}
\pagestyle{myheadings}

\itwmaketitle
\thispagestyle{myheadings}

\footnotetext[1]{This work was supported in part by the Natural Science Foundation of China under Grant NSFC-60472079 and by the Chinese Specialized Research Fund for the Doctoral Program of Higher Education under Grant 2004-0335099.}

\begin{itwabstract}
An enhanced covering lemma for a Markov chain is proved in this paper, and then the distributed source coding problem of correlated general sources with one average distortion criterion under fixed-length coding is investigated. Based on the enhanced lemma, a sufficient and necessary condition for determining the achievability of rate-distortion triples is given.
\end{itwabstract}

\begin{itwpaper}

\itwsection{Introduction}

In the classic problem of multiterminal source coding, $M$ ($M \ge 2$) correlated general sources have to be compressed separately from each other in a lossy fashion, i.e., with respect to a fidelity criterion, and then decoded by the common decoder which has access to a side information source that is correlated with the sources to be compressed. This situation is illustrated in Fig. \ref{fig:Problem}, and it is also called distributed source coding.
\begin{figure}[htbp]
\centering
\includegraphics[width=3in]{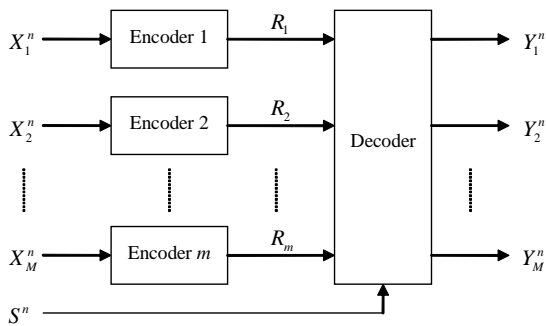}
\caption{Separate compression of $M$ correlated general sources with side information at the decoder}
\label{fig:Problem}
\end{figure}
The well-known Slepian-Wolf coding problem and the Wyner-Ziv coding problem can be regarded as two special cases of this situation. These two special cases were solved in 1970's for stationary memoryless sources \cite{MSC:Slepian197307, MSC:Wyner197601}, and later extended to the case of general sources \cite{MSC:Miyake199509, MSC:Iwata200206}. However, for this general problem, no conclusive results are available to date. Even for the special case that the sources are memoryless and stationary and the distortion measure is additive, only inner and outer bounds are derived in \cite{MSC:Berger197707, MSC:Gastpar200411}, etc. Recently, in \cite{MSC:Yang200604}, we adopt an information-spectrum approach to solve this open problem for general sources with maximum distortion criterions under fixed-length coding and a general formula for the rate-distortion region is obtained. Though the formula in \cite{MSC:Yang200604} is incomputable in general and can not be used to obtain the single letter rate-distortion region for correlated memoryless sources, it does provide a very general sufficient condition, which includes many previous results (e.g. in \cite{MSC:Berger197707,MSC:Iwata200206}) as its special cases.

In this paper, we goes further to investigate the problem with average distortion criterions under fixed-length coding. Since the covering lemma for a Markov chain plays an important role in the proofs of these kinds of problems, we established an enhanced covering lemma in Section \ref{sec:CoveringLemma}, which is the main contribution of this paper. Then in Section \ref{sec:MultiterminalSourceCoding}, we investigate the distributed source coding problem of two correlated general sources with one average distortion criterion under fixed-length coding.

\itwsection{Covering Lemma}\label{sec:CoveringLemma}

In this section, we will prove a covering lemma in a very general form. For comparison, the original covering lemma in \cite{MSC:Iwata200206} is first shown below.
\begin{lemma}[Lemma 1 in \cite{MSC:Iwata200206}]\label{le:Lemma1}
Let $U^n$, $V^n$ and $W^n$ be random variables which take values in finite sets $\mathcal{U}^n$, $\mathcal{V}^n$ and $\mathcal{W}^n$, respectively, and satisfy a Markov condition
$$
P_{U^nV^nW^n} = P_{W^nV^n}P_{U^n|V^n}
$$
for each $n$. Now let $\{\Psi_n\}_{n=1}^\infty$ be a sequence of arbitrary functions such that
\begin{IEEEeqnarray*}{c}
\Psi_n: \mathcal{U}^n \times \mathcal{W}^n \to \{0, 1\},\\
\lim_{n \to \infty} \Pr\{\Psi_n(U^n, W^n) = 1\} = 0.
\end{IEEEeqnarray*}
Then, for any $\gamma > 0$, there exists a sequence $\{F_n\}_{n=1}^\infty$ of function $F_n: \mathcal{V}^n \to \{\bm{w}_i\}_{i=1}^{M_n'} \cap \mathcal{W}^n$ such that
\begin{IEEEeqnarray*}{c}
M_n' = \ceil{e^{n(\overline{I}(\bm{V}; \bm{W}) + \gamma)}},\\
\lim_{n \to \infty} \Pr\{\Psi_n(U^n, F_n(V^n)) = 1\} = 0.
\end{IEEEeqnarray*}
\end{lemma}
The quantity $\overline{I}(\bm{V}; \bm{W})$ in Lemma \ref{le:Lemma1} is called the spectral sup-mutual information rate in the methods of information-spectrum \cite{MSC:Han200300}, and it is defined by
\begin{equation}\label{eq:DefinitionSupI}
\overline{I}(\bm{V}; \bm{W}) \eqdef \plimsup_{n \to \infty} \frac{1}{n} \ln \frac{P_{V^nW^n}(V^n, W^n)}{P_{V^n}(V^n) P_{W^n}(W^n)},
\end{equation}
where $\bm{V}$ and $\bm{W}$ denote the random sequences $\{V^n\}_{n=1}^\infty$ and $\{W^n\}_{n=1}^\infty$ which are called general sources in the methods of information-spectrum, and
$$
\plimsup_{n \to \infty} Z_n \eqdef \inf\biggl\{\alpha \bigg| \lim_{n \to \infty} \Pr\{Z_n > \alpha\} = 0 \biggr\}
$$
denotes the limit superior in probability of the sequence $\{Z_n\}_{n=1}^\infty$ of real-valued random variables.

Now, let us establish the enhanced covering lemma.
\begin{lemma}\label{le:EnhancedCoveringLemma}
Let $U_n$, $V_n$ and $W_n$ be random variables which take values in finite sets $\mathcal{U}_n$, $\mathcal{V}_n$ and $\mathcal{W}_n$, respectively, and satisfy a Markov condition
$$
P_{U_nV_nW_n} = P_{U_nV_n}P_{W_n|V_n}
$$
for each $n$. Now let $\{A_n\}_{n=1}^\infty$ be a sequence of arbitrary sets in $\mathcal{U}_n \times \mathcal{W}_n$ satisfying
\begin{equation}\label{eq:ConditionA}
\lim_{n \to \infty} \Pr\{(U_n, W_n) \in A_n\} = 1,
\end{equation}
and let $\{d_n\}_{n=1}^\infty$ be a sequence of arbitrary functions $d_n: \mathcal{U}_n \times \mathcal{W}_n \to [0, \infty)$ satisfying
\begin{equation}\label{eq:BoundOfDistortion}
D_0 \eqdef \sup_{n \ge 1} \max \{\mathrm{Im}\:d_n\} < \infty,
\end{equation}
where $\mathrm{Im}\:d_n$ denotes the image of $d_n$, then for any $\gamma > 0$, there exits a sequence $\{F_n\}_{n=1}^\infty$ of random functions $F_n: \mathcal{V}_n \to \mathcal{W}_n$ such that
\begin{IEEEeqnarray}{c}
|F_n(\mathcal{V}_n)| \le \ceil{e^{n(\overline{I}(\bm{V}; \bm{W}) + \gamma)}}, \label{eq:CardinalityF} \\
\lim_{n \to \infty} \Pr\{(U_n, F_n(V_n)) \in A_n\} = 1, \label{eq:CoveringA} \\
\limsup_{n \to \infty} \bigl( E[d_n(U_n, F_n(V_n))] - E[d_n(U_n, W_n)] \bigr) \le 0, \label{eq:Coveringd}
\end{IEEEeqnarray}
where $\bm{V} = \{V_n\}_{n=1}^\infty$, $\bm{W}= \{W_n\}_{n=1}^\infty$.
\end{lemma}

\begin{proof}
Let us define
\begin{IEEEeqnarray}{rCl}
\eta_n^{(1)}(\bm{v}, \bm{w}) &\eqdef &\sum_{\bm{u} \in \mathcal{U}_n} P_{U_n|V_nW_n}(\bm{u}|\bm{v},\bm{w}) 1\{(\bm{u}, \bm{w}) \not \in A_n\} \IEEEnonumber \\
&= &\sum_{\bm{u} \in \mathcal{U}_n} P_{U_n|V_n}(\bm{u}|\bm{v}) 1\{(\bm{u}, \bm{w}) \not \in A_n\} \label{eq:DefinitionEta1}
\end{IEEEeqnarray}
and
\begin{IEEEeqnarray}{rCl}
\eta_n^{(2)}(\bm{v}, \bm{w}) &\eqdef &\sum_{\bm{u} \in \mathcal{U}_n} P_{U_n|V_nW_n}(\bm{u}|\bm{v},\bm{w}) d_n(\bm{u}, \bm{w}) \IEEEnonumber \\
&= &\sum_{\bm{u} \in \mathcal{U}_n} P_{U_n|V_n}(\bm{u}|\bm{v}) d_n(\bm{u}, \bm{w}) \label{eq:DefinitionEta2}
\end{IEEEeqnarray}
for $\bm{v} \in \mathcal{V}_n$, $\bm{w} \in \mathcal{W}_n$. Then, it follows from \eqref{eq:ConditionA} and \eqref{eq:BoundOfDistortion} that
\begin{IEEEeqnarray*}{c}
\lim_{n \to \infty} E[\eta_n^{(1)}(V_n, W_n)] = 0, \\
\max_{\bm{v} \in \mathcal{V}_n, \bm{w} \in \mathcal{W}_n} \eta_n^{(2)}(\bm{v}, \bm{w}) \le D_0, \\
E[\eta_n^{(2)}(V_n, W_n)] = E[d_n(U_n, W_n)].
\end{IEEEeqnarray*}
We denote $E[\eta_n^{(1)}(V_n, W_n)]$ by $\delta_n$, and define the set
\begin{equation}\label{eq:DefinitionT1}
T_n^{(1)} \eqdef \bigl\{(\bm{v}, \bm{w}) \in \mathcal{V}_n \times \mathcal{W}_n \big| \eta_n^{(1)}(\bm{v}, \bm{w}) \le \sqrt{\delta_n} \bigr\}.
\end{equation}
Clearly, by Markov's inequality, we have
\begin{equation}\label{eq:InequalityT1}
\Pr\{(V_n, W_n) \not \in T_n^{(1)}\} \le \frac{E[\eta_n^{(1)}(V_n, W_n)]}{\sqrt{\delta_n}} = \sqrt{\delta_n}.
\end{equation}
Letting $\rho$ be an arbitrary nonnegative real numbers, we define
\begin{equation}\label{eq:DefinitionT2}
T_n^{(2)}(\rho) \eqdef \biggl\{ (\bm{v}, \bm{w}) \in \mathcal{V}_n \times \mathcal{W}_n \bigg| \frac{1}{n} \ln \frac{P_{W_n|V_n}(\bm{w}|\bm{v})}{P_{W_n}(\bm{w})} \le \rho \biggr\}.
\end{equation}

Next, set
\begin{equation}\label{eq:DefinitionM}
M_n = \ceil{e^{n(\overline{I}(\bm{V}; \bm{W}) + \gamma)}}.
\end{equation}
We generate a sequence $W_n^{M_n} = \{W_{n,i}\}_{i=1}^{M_n}$, each independently subject to the identical probability distribution $P_{W_n}$. Now, let us define the random function $F_n: \mathcal{V}_n \to \mathcal{W}_n$ with respect to the random sequence $W_n^{M_n}$. For any $\bm{v} \in \mathcal{V}_n$, define
\begin{subnumcases}{F_n(\bm{v}) \eqdef\label{eq:DefinitionF}}
\bm{w}_{\min}(\bm{v}), & $S_n(\bm{v}, W_n^{M_n}) \neq \emptyset$ \\
\bm{w}'_{\min}(\bm{v}), & $S_n(\bm{v}, W_n^{M_n}) = \emptyset$
\end{subnumcases}
where
\begin{IEEEeqnarray*}{rCl}
\bm{w}_{\min}(\bm{v}) &\eqdef &\arg \min_{\bm{w} \in S_n(\bm{v}, W_n^{M_n})} \eta_n^{(2)}(\bm{v}, \bm{w}), \\
\bm{w}'_{\min}(\bm{v}) &\eqdef &\arg \min_{\bm{w} \in \{\bm{w}_i\}_{i=1}^{M_n}} \eta_n^{(2)}(\bm{v}, \bm{w}), \\
S_n(\bm{v}, \bm{w}^{M_n}) &\eqdef &\{\bm{w}_i\}_{i=1}^{M_n} \cap \{\bm{w} \in \mathcal{W}_n | (\bm{v}, \bm{w}) \in T_n^{(1)} \}.
\end{IEEEeqnarray*}
Clearly, $F_n$ satisfies the requirement \eqref{eq:CardinalityF}. Next, let us estimate the upper bound of $\Pr\{(U_n, F_n(V_n)) \not \in A_n\}$ and $E[d_n(U_n, F_n(V_n))]$. First, we have
\begin{IEEEeqnarray*}{Cl}
&\Pr\{(U_n, F_n(V_n)) \not \in A_n\} \\
= &\sum_{\bm{v} \in \mathcal{V}_n} P_{V_n}(\bm{v}) \sum_{\bm{w}^{M_n} \in \mathcal{W}_n^{M_n}} P_{W_n^{M_n}}(\bm{w}^{M_n}) \\
&\sum_{\bm{u} \in \mathcal{U}_n} P_{U_n|V_n}(\bm{u}|\bm{v}) 1\{(\bm{u}, f_n(\bm{v})) \not \in A_n\} \\
\eqvar{(a)} &\sum_{\bm{v} \in \mathcal{V}_n} P_{V_n}(\bm{v}) \sum_{\bm{w}^{M_n} \in \mathcal{W}_n^{M_n}} P_{W_n^{M_n}}(\bm{w}^{M_n}) \eta_n^{(1)}(\bm{v}, f_n(\bm{v})) \\
= &\sum_{\bm{v} \in \mathcal{V}_n} P_{V_n}(\bm{v}) \sum_{\bm{w}^{M_n} \in \mathcal{W}_n^{M_n}} P_{W_n^{M_n}}(\bm{w}^{M_n}) \eta_n^{(1)}(\bm{v}, f_n(\bm{v})) \\
&\bigl( 1\{(\bm{v}, f_n(\bm{v})) \in T_n^{(1)}\} + 1\{(\bm{v}, f_n(\bm{v})) \not \in T_n^{(1)}\} \bigr) \\
\levar{(b)} &\sqrt{\delta_n} + \sum_{\bm{v} \in \mathcal{V}_n} P_{V_n}(\bm{v}) \\
&\sum_{\bm{w}^{M_n} \in \mathcal{W}_n^{M_n}} P_{W_n^{M_n}}(\bm{w}^{M_n}) 1\{(\bm{v}, f_n(\bm{v})) \not \in T_n^{(1)}\} \\
= &\sqrt{\delta_n} + \sum_{\bm{v} \in \mathcal{V}_n} P_{V_n}(\bm{v}) \Pr\{(\bm{v}, F_n(\bm{v})) \not \in T_n^{(1)}\},
\end{IEEEeqnarray*}
where (a) follows from \eqref{eq:DefinitionEta1}, and (b) follows from \eqref{eq:DefinitionT1}. Furthermore, we have
\begin{IEEEeqnarray*}{Cl}
&\Pr\{(\bm{v}, F_n(\bm{v})) \not \in T_n^{(1)}\} \\
\eqvar{(a)} &\sum_{\bm{w}^{M_n} \in \mathcal{W}_n^{M_n}} P_{W_n^{M_n}}(\bm{w}^{M_n}) 1\{S_n(\bm{v}, \bm{w}^{M_n}) = \emptyset\} \\
\eqvar{(b)} &\sum_{\bm{w}^{M_n} \in \mathcal{W}_n^{M_n}} \prod_{i=1}^{M_n} P_{W_n}(\bm{w}_i) 1\{(\bm{v}, \bm{w}_i) \not \in T_n^{(1)}\} \\
= &\biggl( \sum_{\bm{w} \in \mathcal{W}_n} P_{W_n}(\bm{w}) 1\{(\bm{v}, \bm{w}) \not \in T_n^{(1)}\} \biggr)^{M_n} \\
= &\biggl( 1 - \sum_{\bm{w} \in \mathcal{W}_n} P_{W_n}(\bm{w}) 1\{(\bm{v}, \bm{w}) \in T_n^{(1)}\} \biggr)^{M_n} \\
\le &\biggl( 1 - \sum_{\bm{w} \in \mathcal{W}_n} P_{W_n}(\bm{w}) 1\{(\bm{v}, \bm{w}) \in T_n^{(1)} \cap T_n^{(2)}(\rho)\} \biggr)^{M_n} \\
\levar{(c)} &\biggl( 1 - e^{-n\rho} \sum_{\bm{w} \in \mathcal{W}_n} P_{W_n|V_n}(\bm{w}|\bm{v}) \\
&1\{(\bm{v}, \bm{w}) \in T_n^{(1)} \cap T_n^{(2)}(\rho)\} \biggr)^{M_n} \\
\levar{(d)} &1 - \sum_{\bm{w} \in \mathcal{W}_n} P_{W_n|V_n}(\bm{w}|\bm{v}) \\
&1\{(\bm{v}, \bm{w}) \in T_n^{(1)} \cap T_n^{(2)}(\rho)\} + e^{-M_ne^{-n\rho}},
\end{IEEEeqnarray*}
where (a) and (b) follows from \eqref{eq:DefinitionF}, (c) follows from \eqref{eq:DefinitionT2}, and (d) follows from the inequality $(1 - xy)^n \le 1 - x + e^{-yn}$ for $0 \le x, y \le 1$, $n \ge 1$. Then we have
\begin{IEEEeqnarray*}{Cl}
&\Pr\{(U_n, F_n(V_n)) \not \in A_n\} \\
= &\sqrt{\delta_n} + \sum_{\bm{v} \in \mathcal{V}_n} P_{V_n}(\bm{v}) \biggl( 1 - \sum_{\bm{w} \in \mathcal{W}_n} P_{W_n|V_n}(\bm{w}|\bm{v}) \\
&1\{(\bm{v}, \bm{w}) \in T_n^{(1)} \cap T_n^{(2)}(\rho)\} + e^{-M_ne^{-n\rho}} \biggr) \\
= &\sqrt{\delta_n} + \Pr\{(V_n, W_n) \not \in T_n^{(1)} \cap T_n^{(2)}(\rho)\} + e^{-M_ne^{-n\rho}} \\
\le &\sqrt{\delta_n} + \Pr\{(V_n, W_n) \not \in T_n^{(1)}\} \\
&+\: \Pr\{(V_n, W_n) \not \in T_n^{(2)}(\rho)\} + e^{-M_ne^{-n\rho}} \\
\levar{(a)} &2\sqrt{\delta_n} + \Pr\{(V_n, W_n) \not \in T_n^{(2)}(\rho)\} + e^{-M_ne^{-n\rho}},
\end{IEEEeqnarray*}
where (a) follows from \eqref{eq:InequalityT1}. Letting $\rho = \overline{I}(\bm{V}; \bm{W}) + \frac{\gamma}{2}$, we have
\begin{IEEEeqnarray*}{Cl}
&\Pr\{(U_n, F_n(V_n)) \not \in A_n\} \\
\levar{(a)} &2\sqrt{\delta_n} + \Pr\{(V_n, W_n) \not \in T_n^{(2)}(\overline{I}(\bm{V}; \bm{W}) + \frac{\gamma}{2})\} \\
&+\: e^{-e^{\frac{n\gamma}{2}}} \\
\relvar{\to}{(b)} &0
\end{IEEEeqnarray*}
as $n \to \infty$, where (a) follows from \eqref{eq:DefinitionM}, and (b) follows from \eqref{eq:DefinitionSupI} and \eqref{eq:DefinitionT2}. This concludes \eqref{eq:CoveringA}.

Second, the expectation $E[d_n(U_n, F_n(V_n))]$ can be written as
\begin{IEEEeqnarray*}{Cl}
&E[d_n(U_n, F_n(V_n))] \\
= &\sum_{\bm{v} \in \mathcal{V}_n} P_{V_n}(\bm{v}) \sum_{\bm{w}^{M_n} \in \mathcal{W}_n^{M_n}} P_{W_n^{M_n}}(\bm{w}^{M_n}) \\
&\sum_{\bm{u} \in \mathcal{U}_n} P_{U_n|V_n}(\bm{u}|\bm{v}) d_n(\bm{u}, f_n(\bm{v})) \\
\eqvar{(a)} &\sum_{\bm{v} \in \mathcal{V}_n} P_{V_n}(\bm{v}) \sum_{\bm{w}^{M_n} \in \mathcal{W}_n^{M_n}} P_{W_n^{M_n}}(\bm{w}^{M_n}) \\
&\int_0^{D_0} 1\{ \eta_n^{(2)}(\bm{v}, f_n(\bm{v})) \ge \beta \} d\beta \\
= &\sum_{\bm{v} \in \mathcal{V}_n} P_{V_n}(\bm{v}) \int_0^{D_0} Q(\bm{v}, \beta) d\beta,
\end{IEEEeqnarray*}
where (a) follows from \eqref{eq:DefinitionEta2}, and
$$
Q(\bm{v}, \beta) \eqdef \sum_{\bm{w}^{M_n} \in \mathcal{W}_n^{M_n}} P_{W_n^{M_n}}(\bm{w}^{M_n}) 1\{ \eta_n^{(2)}(\bm{v}, f_n(\bm{v})) \ge \beta \}.
$$
Furthermore, we have
\begin{IEEEeqnarray*}{Cl}
&Q(\bm{v}, \beta) \\
\levar{(a)} &\sum_{\bm{w}^{M_n} \in \mathcal{W}_n^{M_n}} \prod_{i=1}^{M_n} P_{W_n}(\bm{w}_i) 1\{ \eta_n^{(2)}(\bm{v}, \bm{w}_i) \ge \beta \\
&\mbox{ or } (\bm{v}, \bm{w}_i) \not \in T_n^{(1)} \} \\
= &\biggl( \sum_{\bm{w} \in \mathcal{W}_n} \!\!\!\! P_{W_n}(\bm{w}) 1\{ \eta_n^{(2)}(\bm{v}, \bm{w}) \ge \beta \! \mbox{ or } \! (\bm{v}, \bm{w}) \not \in T_n^{(1)} \} \biggr)^{M_n} \\
= &\biggl( \! 1 - \!\!\! \sum_{\bm{w} \in \mathcal{W}_n} \!\!\!\! P_{W_n}(\bm{w}) 1\{ \eta_n^{(2)}(\bm{v}, \bm{w}) < \beta, (\bm{v}, \bm{w}) \in T_n^{(1)} \} \! \biggr)^{M_n} \\
\le &\biggl( 1 - \sum_{\bm{w} \in \mathcal{W}_n} P_{W_n}(\bm{w}) 1\{ \eta_n^{(2)}(\bm{v}, \bm{w}) < \beta, \\
&(\bm{v}, \bm{w}) \in T_n^{(1)} \cap T_n^{(2)}(\rho) \} \biggr)^{M_n} \\
\levar{(b)} &\biggl( 1 - e^{-n\rho} \sum_{\bm{w} \in \mathcal{W}_n} P_{W_n|V_n}(\bm{w}|\bm{v}) 1\{ \eta_n^{(2)}(\bm{v}, \bm{w}) < \beta, \\
&(\bm{v}, \bm{w}) \in T_n^{(1)} \cap T_n^{(2)}(\rho) \} \biggr)^{M_n} \\
\levar{(c)} &1 - \sum_{\bm{w} \in \mathcal{W}_n} P_{W_n|V_n}(\bm{w}|\bm{v}) 1\{ \eta_n^{(2)}(\bm{v}, \bm{w}) < \beta, \\
&(\bm{v}, \bm{w}) \in T_n^{(1)} \cap T_n^{(2)}(\rho) \} + e^{-M_ne^{-n\rho}} \\
= &\sum_{\bm{w} \in \mathcal{W}_n} P_{W_n|V_n}(\bm{w}|\bm{v}) 1\{ \eta_n^{(2)}(\bm{v}, \bm{w}) \ge \beta \mbox{ or } (\bm{v}, \bm{w}) \not \in T_n^{(1)} \\
&\mbox{ or } (\bm{v}, \bm{w}) \not \in T_n^{(2)}(\rho) \} + e^{-M_ne^{-n\rho}}
\end{IEEEeqnarray*}
where (a) follows from \eqref{eq:DefinitionF}, (b) follows from \eqref{eq:DefinitionT2}, and (c) also follows from the inequality $(1 - xy)^n \le 1 - x + e^{-yn}$ for $0 \le x, y \le 1$, $n \ge 1$. Then we have
\begin{IEEEeqnarray*}{Cl}
&E[d_n(U_n, F_n(V_n))] \\
\le &\sum_{\bm{v} \in \mathcal{V}_n} P_{V_n}(\bm{v}) \int_0^{D_0} \biggl( \sum_{\bm{w} \in \mathcal{W}_n} P_{W_n|V_n}(\bm{w}|\bm{v}) \\
&1\{ \eta_n^{(2)}(\bm{v}, \bm{w}) \ge \beta \mbox{ or } (\bm{v}, \bm{w}) \not \in T_n^{(1)} \\
&\mbox{ or } (\bm{v}, \bm{w}) \not \in T_n^{(2)}(\rho) \} + e^{-M_ne^{-n\rho}} \biggr) d\beta \\
\le &\sum_{\bm{v} \in \mathcal{V}_n, \bm{w} \in \mathcal{W}_n} P_{V_nW_n}(\bm{v}, \bm{w}) \int_0^{D_0} \biggl( 1\{\eta_n^{(2)}(\bm{v}, \bm{w}) \ge \beta\} \\
&+\: 1\{(\bm{v}, \bm{w}) \not \in T_n^{(1)}\} + 1\{(\bm{v}, \bm{w}) \not \in T_n^{(2)}(\rho)\} \biggr) d\beta \\
&+\: D_0 e^{-M_ne^{-n\rho}} \\
\le &\sum_{\bm{v} \in \mathcal{V}_n, \bm{w} \in \mathcal{W}_n} P_{V_nW_n}(\bm{v}, \bm{w}) \biggl( \eta_n^{(2)}(\bm{v}, \bm{w}) \\
&+\: D_0 1\{(\bm{v}, \bm{w}) \not \in T_n^{(1)}\} + D_0 1\{(\bm{v}, \bm{w}) \not \in T_n^{(2)}(\rho)\} \biggr) \\
&+\: D_0 e^{-M_ne^{-n\rho}} \\
\eqvar{(a)} &E[d_n(U_n, W_n)] + D_0 \Pr\{(V_n, W_n) \not \in T_n^{(1)}\} \\
&+\: D_0 \Pr\{(V_n, W_n) \not \in T_n^{(2)}(\rho)\} + D_0 e^{-M_ne^{-n\rho}}
\end{IEEEeqnarray*}
where (a) follows from \eqref{eq:DefinitionEta2}. Letting $\rho = \overline{I}(\bm{V}; \bm{W}) + \frac{\gamma}{2}$, we have
\begin{IEEEeqnarray*}{Cl}
&E[d_n(U_n, F_n(V_n))] - E[d_n(U_n, W_n)] \\
\levar{(a)} &D_0 \sqrt{\delta_n} + D_0 \Pr\{(V_n, W_n) \not \in T_n^{(2)}(\overline{I}(\bm{V}; \bm{W}) + \frac{\gamma}{2})\} \\
&+\: D_0 e^{-e^{\frac{n\gamma}{2}}} \\
\relvar{\to}{(b)} &0
\end{IEEEeqnarray*}
as $n \to \infty$, where (a) follows from \eqref{eq:InequalityT1} and \eqref{eq:DefinitionM}, and (b) follows from \eqref{eq:DefinitionSupI} and \eqref{eq:DefinitionT2}. This concludes \eqref{eq:Coveringd} and hence completes the proof.
\end{proof}

\begin{remark}
The main idea of this proof is a combination of the ideas in the proofs of Lemma 1 in \textnormal{\cite{MSC:Iwata200206}} and Theorem 5.5.1 in \textnormal{\cite{MSC:Han200300}}. However, such a method has its own limitation. Because the minimum operation in \eqref{eq:DefinitionF} should be applied to an ordered set, we can establish covering lemmas with only one average distortion criterion.
\end{remark}

By Lemma \ref{le:EnhancedCoveringLemma}, we can easily obtain the following proposition that is also a generalized version of Lemma 1 in \textnormal{\cite{MSC:Iwata200206}}.

\begin{proposition}[Lemma 2 in \cite{MSC:Yang200604}]
Let $U_n$, $V_n$ and $W_n$ be random variables which take values in finite sets $\mathcal{U}_n$, $\mathcal{V}_n$ and $\mathcal{W}_n$, respectively, and satisfy a Markov condition
$$
P_{U_nV_nW_n} = P_{U_nV_n}P_{W_n|V_n}
$$
for each $n$. Now let $\{B_n\}_{n=1}^\infty$ be a sequence of arbitrary sets in $\mathcal{U}_n \times \mathcal{W}_n$ satisfying
\begin{equation}
\liminf_{n \to \infty} \Pr\{(U_n, W_n) \in B_n\} = \epsilon,
\end{equation}
then for any $\gamma > 0$, there exits a sequence $\{F_n\}_{n=1}^\infty$ of random functions $F_n: \mathcal{V}_n \to \mathcal{W}_n$ such that
\begin{IEEEeqnarray}{c}
|F_n(\mathcal{V}_n)| \le \ceil{e^{n(\overline{I}(\bm{V}; \bm{W}) + \gamma)}}, \\
\liminf_{n \to \infty} \Pr\{(U_n, F_n(V_n)) \in B_n\} \ge \epsilon.
\end{IEEEeqnarray}
\end{proposition}

\begin{proof}
Letting $A_n = \mathcal{U}_n \times \mathcal{W}_n$ and $d_n = 1\{(\bm{u}, \bm{w}) \not \in B_n\}$ and then applying Lemma \ref{le:EnhancedCoveringLemma}, we have
$$
\limsup_{n \to \infty} (E[d_n(U_n, F_n(V_n)] - E[d_n(U_n, W_n)]) \le 0,
$$
where $F_n$ is the random function constructed in the proof of Lemma \ref{le:EnhancedCoveringLemma}. Then we have
\begin{IEEEeqnarray*}{Cl}
&\liminf_{n \to \infty} \Pr\{(U_n, F_n(V_n)) \in B_n\} \\
= &1 - \limsup_{n \to \infty} E[d_n(U_n, F_n(V_n))] \\
\ge &1 - \bigl[ \limsup_{n \to \infty} E[d_n(U_n, W_n)] \\
&+\: \limsup_{n \to \infty} (E[d_n(U_n, F_n(V_n))] - E[d_n(U_n, W_n)]) \bigr] \\
\ge &1 - \limsup_{n \to \infty} E[d_n(U_n, W_n)] \\
= &\liminf_{n \to \infty} \Pr\{(U_n, W_n) \in B_n\}.
\end{IEEEeqnarray*}
This proves the proposition.
\end{proof}

\itwsection{Multiterminal Source Coding}\label{sec:MultiterminalSourceCoding}

In this section, we will investigate the sufficient and necessary condition for multiterminal source coding with one average distortion criterion.

As depicted in Figure \ref{fig:Problem}, a multiterminal source coding system can be stated as follows. For simplicity, we will only consider the case of two terminals without side information at the decoder. Given a pair of correlated general sources $(\bm{X}_1, \bm{X}_2)$ with finite alphabet $\mathcal{X}_1 \times \mathcal{X}_2$, each characterized by an infinite sequence
$$
\bm{X}_m = \{X_m^n = (X_{m,1}^{(n)}, X_{m,2}^{(n)}, \cdots, X_{m,n}^{(n)})\}_{n=1}^\infty, \quad m = 1, 2
$$
of $n$-dimensional random variables $X_m^n$ taking values in the $n$-th Cartesian product $\mathcal{X}_m^n$, the $n$-length source outputs $(X_1^n, X_2^n)$ are separately encoded into a pair of fixed-length codewords $(\phi_n^{(1)}(X_1^n), \phi_n^{(2)}(X_2^n))$, and then the common decoder observes these codewords to reproduce the estimates $(Y_1^n, Y_2^n) = (\psi_n^{(1)}(X_1^n, X_2^n), \psi_n^{(2)}(X_1^n, X_2^n))$ of $(X_1^n, X_2^n)$. Here, the pair of encoders are maps defined by
$$
\phi_n^{(m)}: \mathcal{X}_m^n \to \mathcal{I}_{L_n^{(m)}} \eqdef \{1, 2, \cdots, L_n^{(m)}\},
$$
and the rate of each encoder is calculated by
$$
R(\phi_n^{(m)}) \eqdef \frac{\ln |\phi_n^{(m)}(\mathcal{X}_m^n)|}{n}.
$$
The decoders $\psi_n^{(m)}$ ($m = 1, 2$) are maps defined by
$$
\psi_n^{(m)}: \mathcal{I}_{L_n^{(1)}} \times \mathcal{I}_{L_n^{(2)}} \to \mathcal{Y}_m^n,
$$
where $\mathcal{Y}_m^n$ denotes the $n$-th Cartesian product space in which the $n$-dimensional estimate $Y_m^n$ takes values. Next, let us define the distortion measure. A general distortion measure $\bm{d}$ is a sequence $\{d_n\}_{n=1}^\infty$ of functions $d_n: \mathcal{X}_1^n \times \mathcal{X}_2^n \times \mathcal{Y}_1^n \times \mathcal{Y}_2^n \to [0, +\infty)$, and hence the average distortion is calculated by
$$
E[d_n(X_1^n, X_2^n, Y_1^n, Y_2^n)].
$$
Then for such a system with an average distortion criterion, a rate-distortion triple $(R_1, R_2, D)$ is said to be $fa$-achievable if and only if there exists a sequence of fixed-length codes such that
\begin{IEEEeqnarray*}{l}
\limsup_{n \to \infty} R(\phi_n^{(1)}) \le R_1, \quad \limsup_{n \to \infty} R(\phi_n^{(2)}) \le R_2, \\
\limsup_{n \to \infty} E \bigl[ d_n( X_1^n, X_2^n, Y_1^n, Y_2^n) \bigr] \le D.
\end{IEEEeqnarray*}
Therefore, our task is to find the set of all the triples $(R_1, R_2, D)$ that are $fa$-achievable. By applying Lemma \ref{le:EnhancedCoveringLemma}, we proved the following sufficient and necessary condition.

\begin{theorem}\label{th:MSCfaVersion}
For a pair of correlated general sources $(\bm{X}_1, \bm{X}_2)$ and a distortion measure $\bm{d}$ satisfying the condition \eqref{eq:BoundOfDistortion}, the rate-distortion triple $(R_1, R_2, D)$ is $fa$-achievable if and only if there exist a pair of random sequences $(\bm{Z}^{(1)}, \bm{Z}^{(2)}) = \{(Z_n^{(1)}, Z_n^{(2)})\}_{n=1}^\infty$ with the alphabet $\{\mathcal{Z}_n^{(1)} \times \mathcal{Z}_n^{(2)}\}_{n=1}^\infty$ and a pair of function sequences $(\bm{h}^{(1)}, \bm{h}^{(2)}) = \{(h_n^{(1)}, h_n^{(2)})\}_{n=1}^\infty$ defined by
$$
h_n^{(m)}: \mathcal{Z}_n^{(1)} \times \mathcal{Z}_n^{(2)} \to \mathcal{Y}_m^n
$$
such that
\begin{IEEEeqnarray*}{l}
P_{X_1^n X_2^n Z_n^{(1)} Z_n^{(2)}} = P_{X_1^n X_2^n} P_{Z_n^{(1)} | X_1^n} P_{Z_n^{(2)} | X_2^n}, \quad \forall n \ge 1 \\
\limsup_{n \to \infty} E \biggl[ d_n \bigl( X_1^n, X_2^n, h_n^{(1)}(Z_n^{(1)}, Z_n^{(2)}), \\
\qquad h_n^{(2)}(Z_n^{(1)}, Z_n^{(2)}) \bigr) \biggr] \le D,
\end{IEEEeqnarray*}
and
\begin{IEEEeqnarray*}{rCl}
R_1 &\ge &\overline{I}(\bm{X}_1; \bm{Z}^{(1)}) - \underline{I}(\bm{Z}^{(1)}; \bm{Z}^{(2)}), \\
R_2 &\ge &\overline{I}(\bm{X}_2; \bm{Z}^{(2)}) - \underline{I}(\bm{Z}^{(1)}; \bm{Z}^{(2)}), \\
R_1 + R_2 &\ge &\overline{I}(\bm{X}_1; \bm{Z}^{(1)}) + \overline{I}(\bm{X}_2; \bm{Z}^{(2)}) - \underline{I}(\bm{Z}^{(1)}; \bm{Z}^{(2)}).
\end{IEEEeqnarray*}
\end{theorem}

\begin{remark}
In Theorem \ref{th:MSCfaVersion}, the quantity $\underline{I}(\bm{X}; \bm{Y})$ denotes the spectral inf-mutual information rate defined by
$$
\pliminf_{n \to \infty} \frac{1}{n} \ln \frac{P_{X_nY_n}(X_n, Y_n)}{P_{X_n}(X_n) P_{Y_n}(Y_n)},
$$
where
$$
\pliminf_{n \to \infty} Z_n \eqdef - \plimsup_{n \to \infty} (-Z_n).
$$
The proof of Theorem \ref{th:MSCfaVersion} is analogous to the proof of Theorem 1 in \textnormal{\cite{MSC:Yang200604}}, and it relies heavily on Lemma \ref{le:CoveringXSZ} presented below, which is an easy consequence of Lemma \ref{le:EnhancedCoveringLemma}. Note that the rate-distortion function for the Wyner-Ziv problem with an average distortion criterion can be easily obtained by letting $Z_n^{(2)} = X_2^n$ and
$$
R_2 = \overline{H}(\bm{X}_2) \eqdef \plimsup_{n \to \infty} \frac{1}{n} \ln \frac{1}{P_{X_2^n}(X_2^n)}
$$
in Theorem \ref{th:MSCfaVersion}. Furthermore, Theorem \ref{th:MSCfaVersion} can be easily extended to the case of multiple terminals with one average distortion criterion and multiple maximum distortion criterions.
\end{remark}

\begin{lemma}\label{le:CoveringXSZ}
Let $X_1^n$, $X_2^n$, $Z_n^{(1)}$, and $Z_n^{(2)}$ be random variables which take values in finite sets $\mathcal{X}_1^n$, $\mathcal{X}_2^n$, $\mathcal{Z}_n^{(1)}$ and $\mathcal{Z}_n^{(2)}$, respectively, and satisfy
$$
P_{X_1^n X_2^n Z_n^{(1)} Z_n^{(2)}} = P_{X_1^n X_2^n} P_{Z_n^{(1)} | X_1^n} P_{Z_n^{(2)} | X_2^n}
$$
for each $n$. Now let $\{A_n\}_{n=1}^\infty$ be a sequence of arbitrary sets in $\mathcal{X}_1^n \times \mathcal{X}_2^n \times \mathcal{Z}_n^{(1)} \times \mathcal{Z}_n^{(2)}$ satisfying
$$
\lim_{n \to \infty} \Pr\{(X_1^n, X_2^n, Z_n^{(1)}, Z_n^{(2)}) \in A_n\} = 1,
$$
and let $\{d_n\}_{n=1}^\infty$ be a sequence of arbitrary functions $d_n: \mathcal{X}_1^n \times \mathcal{X}_2^n \times \mathcal{Z}_n^{(1)} \times \mathcal{Z}_n^{(2)} \to [0, \infty)$ satisfying the condition \eqref{eq:BoundOfDistortion}, then for any $\gamma_1, \gamma_2 > 0$, there exit two sequences $\{F_n^{(m)}\}_{n=1}^\infty$ of functions $F_n^{(m)}: \mathcal{X}_m^n \to \mathcal{Z}_n^{(m)}$ $(m = 1, 2)$ such that
\begin{IEEEeqnarray*}{c}
|F_n^{(m)}(\mathcal{X}_m^n)| \le \ceil{e^{n(\overline{I}(\bm{X}_m; \bm{Z}^{(m)}) + \gamma_1)}}, \quad m = 1, 2 \\
\lim_{n \to \infty} \Pr\{(X_1^n, X_2^n, F_n^{(1)}(X_1^n), F_n^{(2)}(X_2^n)) \in A_n\} = 1, \\
\limsup_{n \to \infty} \bigl( E[d_n(X_1^n, X_2^n, F_n^{(1)}(X_1^n), F_n^{(2)}(X_2^n))] - \\
\qquad E[d_n(X_1^n, X_2^n, Z_n^{(1)}, Z_n^{(2)})] \bigr) \le 0,
\end{IEEEeqnarray*}
and
$$
\underline{I}(\bm{F}^{(1)}(\bm{X}_1); \bm{F}^{(2)}(\bm{X}_2)) \ge \underline{I}(\bm{Z}^{(1)}; \bm{Z}^{(2)}) - \gamma_2,
$$
where $\bm{F}^{(m)}(\bm{X}_m) \eqdef \{F_n^{(m)}(X_m^n)\}_{n=1}^\infty$.
\end{lemma}

\end{itwpaper}


\begin{itwreferences}

\bibitem{MSC:Slepian197307}
D.~Slepian and J.~K. Wolf, ``Noiseless coding of correlated information
  sources,'' \emph{{IEEE} Trans. Inform. Theory}, vol.~19, no.~4, pp. 471--480,
  July 1973.

\bibitem{MSC:Wyner197601}
A.~D. Wyner and J.~Ziv, ``The rate-distortion function for source coding with
  side information at the decoder,'' \emph{{IEEE} Trans. Inform. Theory},
  vol.~22, no.~1, pp. 1--10, Jan. 1976.

\bibitem{MSC:Miyake199509}
S.~Miyake and F.~Kanaya, ``Coding theorems on correlated general sources,''
  \emph{IEICE Trans. Fundamentals}, vol. E78-A, no.~9, pp. 1063--1070, Sept.
  1995.

\bibitem{MSC:Iwata200206}
K.~Iwata and J.~Muramatsu, ``An information-spectrum approach to
  rate-distortion function with side information,'' \emph{IEICE Trans.
  Fundamentals}, vol. E85-A, no.~6, pp. 1387--1395, June 2002.

\bibitem{MSC:Berger197707}
T.~Berger, ``Multiterminal source coding,'' in \emph{The Information Theory
  Approach to Communications}.\hskip 1em plus 0.5em minus 0.4em\relax New York:
  Springer-Verlag, July 1977, pp. 171--231.

\bibitem{MSC:Gastpar200411}
M.~Gastpar, ``The {Wyner-Ziv} problem with multiple sources,'' \emph{{IEEE}
  Trans. Inform. Theory}, vol.~50, no.~11, pp. 2762--2768, Nov. 2004.

\bibitem{MSC:Yang200604}
S.~Yang and P.~Qiu, ``An information-spectrum approach to multiterminal
  rate-distortion theory,'' submitted to IEEE Trans. Inform. Theory (draft
  available at http://arxiv.org/{\linebreak[0]}abs/cs/0605006).

\bibitem{MSC:Han200300}
T.~S. Han, \emph{Information-Spectrum Methods in Information Theory}.\hskip 1em
  plus 0.5em minus 0.4em\relax Berlin: Springer, 2003.

\end{itwreferences}

\end{document}